\documentstyle[aps,prl,twocolumn,psfig,epsf]{revtex}
\draft
\begin{document}
\noindent
{\bf Comment on ``Electronic fine structure in \\
the electron-hole plasma of SrB$_6$''}
\medskip

In a recent Letter \cite{Pickett}, Rodriguez {\em et al\/}.\
addressed the origin of the high-temperature weak
ferromagnetism found in doped hexaborides \cite{Young} by calculating 
details of the band structure of SrB$_6$ in 
the vicinity of the X-point in the Brillouin zone. They conclude that
crystal field induced electron-hole mixing has an important effect 
on the density of states (DOS) and suggest nesting between electron 
Fermi pockets as an alternative explanation to the excitonic
ferromagnetism \cite{Nature}.
We would like to question several conclusions
in this work \cite{Pickett}.

(i) Rodriguez {\em et al\/}.\ claim that the two bands cross 
only along the symmetry lines $(\xi,0,0)$ and 
$(\frac{\pi}{a},\xi,0)$. This result contradicts a previous 
band structure study, which found band crossing in {\em all 
symmetry planes} passing through the X-point \cite{Hasegawa}.
The latter conclusion has a simple symmetry interpretation.
At the X-point the valence and conduction bands belong to
different irreducible representations of the small group: 
$X_3$, basis function $yz$, and $X_3'$ symmetry, basis function 
$x(y^2-z^2)$, respectively.
The band mixing occurs away from the X-point and
is given by the product $X_3\times X_3'$
\begin{equation}
\Delta_{\rm cryst} \simeq E_b
k_xk_yk_z(k_y^2-k_z^2)/b^5 \ ,
\end{equation}
where $E_b$ is a bandwidth and $b=\frac{2\pi}{a}$ is a reciprocal 
lattice vector. The crystal field hybridization is strongly
suppressed by the high power of momentum in Eq.~(1) 
($k_F\sim 0.1 b$). Therefore, not only is the hybridization more
anisotropic than suggested in Ref.~\cite{Pickett},
but also it is very weak on the scale of the band overlap.
Our TB-LMTO calculations \cite{Nature} as well as FLAPW method 
\cite{Monnier} yield an extremely small band mixing 
for a general direction 
in the Brillouin zone 
in agreement with Eq. (1).

(ii) Even for a strong hybridization, 
the DOS is quite different from the results of Ref.~\cite{Pickett}. 
Fig.~1 shows the DOS for two parabolic bands with anisotropic
effective masses \cite{Pickett,Nature}, band overlap 
$E_G=100$~meV, and two angular dependences for
the mixing matrix element: isotropic, which models
an excitonic gap \cite{Nature}, 
and anisotropic given by Eq.~(1).
The amplitudes were chosen to give the same maximum
splitting
of $2\Delta_{\rm max}=15$~meV.
Because of the anisotropic dispersion the 
DOS is slightly reduced at the Fermi level.
The true gap in the DOS opens only for a sufficiently large
excitonic gap $\Delta>0.16E_G$. 
Note that the tetrahedron method for DOS 
does not allow for crossing of bands, which are numbered in 
increasing order of energy at every k-point, and, therefore, may lead 
to spurious peaks \cite{Pickett} due to artificial anticrossings.

(iii) Correlation effects in the $e$-$h$ plasma 
lead to a forbidden range for the band overlap values in the hexaborides
\cite{PRB}. At the first order metal-insulator transition  
a semiconductor jumps into a dense plasma with
$r_s\approx 1$. Rodriguez {\em et al\/}.\ have estimated 
$r_s=2.5$ for the band overlap $E_G=100$~meV in SrB$_6$,
a value lying
within the forbidden range.
The real overlap must be larger.
Therefore, the conclusion
that 0.5~\% doping
fills hole Fermi pockets in SrB$_6$ \cite{Pickett},
which is
based on the above underestimated
value of $E_G$, 
is by no means reliable and does not exclude
electron-hole pairing at this doping level. 
\vspace*{-17mm}

\begin{figure}
\unitlength1cm
\epsfxsize=6cm
\begin{picture}(2,3)
\put(1,-3){\epsffile{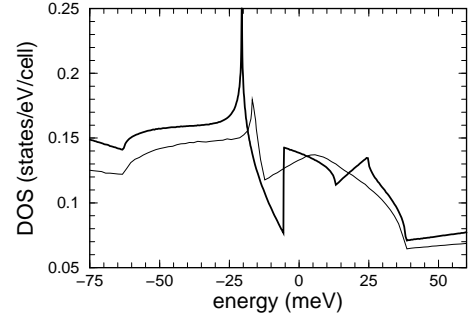}}
\end{picture}
\vspace{3.cm}
\caption{The density of states in SrB$_6$ for two types
of interband hybridization (a) isotropic excitonic gap,
bold line, and (b) anisotropic crystal-field mixing Eq.~(1),
thin line.}
\end{figure}

(iv) Nesting of electron pockets between inequivalent 
X-points was suggested in Ref.~\cite{Pickett} as another
source of CDW and SDW instabilities, which does not involve
holes from the valence band. This suggestion is physically incorrect.
The nesting condition refers to a sign change under translation
in momentum space: $\epsilon(k+Q)\approx-\epsilon(k)$.
Only in such a case can repulsive $e$-$e$ interaction produce
a density-wave instability in $e$-$h$ channel, see e.g. \cite{Fazekas}.

Lastly, the interpretation of the de Haas-van Alphen measurements quoted
in \cite{Pickett} seems to us to be problematic in view
of the high values of the low temperature resistivity
reported for SrB$_6$ \cite{Ott} and its strong dependence
on stoichiometry.
Further information is required 
before a definitive conclusion can be made.

We thank R. Monnier for many useful discussions.

\medskip\noindent
M. E. Zhitomirsky,$^1$ T. M. Rice,$^1$ and V. I. Anisimov$^2$\\
$^1$Theoretische Physik, ETH-Z\"urich, Switzerland \\
$^2$Institute of Metal Physics, Ekaterinburg, Russia


\end{document}